\documentclass[%
reprint,aps,amsmath,amssymb
]{revtex4-2}

\usepackage{color}
\usepackage{graphicx}
\usepackage{dcolumn}
\usepackage{bm}
\usepackage[colorlinks=true,citecolor=blue,urlcolor=blue,linkcolor=blue,hyperindex]{hyperref}
\usepackage{mathtools}
\usepackage{float}
\usepackage{CJK}
\usepackage{booktabs}
\usepackage{subfig}

\makeatother

\begin{document}

	\title{Theory-Guided, Machine-Learning-Accelerated Discovery of a 3D Carbon Nested Nodal-Surface Semimetal}
	
	\author{Shuaihua Zhang}
	\author{Silei Guo}
	\author{Jingxiang Liu}
	\author{Baoxin Hu}
	\author{Yanling Wu}
	\email{ylwu@ysu.edu.cn}
	\author{Jun Li}
	\email{ljcj007@ysu.edu.cn}
	\affiliation{
		State Key Laboratory of Metastable Materials Science and Technology, Hebei Key Laboratory of Microstructural Material Physics, School of Science, Yanshan University, Qinhuangdao, 066004, People’s Republic of China.}
	
	\begin{abstract}
		Extending the Dirac physics of two-dimensional (2D) graphene into three dimensions (3D) carbon allotropes with higher-dimensional band degeneracies remains a central challenge in topological materials science. Here, we propose a general symmetry-engineering principle that systematically transforms graphene's Dirac cone into a 3D nodal surface via controlled layering and registry shift, and employ this principle to guide a machine-learning-accelerated inverse design. By integrating a crystal diffusion variational autoencoder(CDVAE) with a Crystal Transformer, we discover a novel, dynamically and mechanically stable carbon allotrope named \textbf{Netsene} (bct-C$_{24}$), which crystallizes in the body-centered tetragonal \textit{I4/mcm} space group. First-principles calculations confirm that Netsene is a unique nested nodal-surface semimetal: it hosts a complex, double-bowl-shaped nodal-surface system around the Fermi level, protected by non-symmorphic symmetries, alongside Dirac-like linear crossings with Fermi velocities comparable to graphene ($\sim 9 \times 10^5$~m/s). Its non-trivial bulk topology manifests in drumhead surface states, including a nearly flat band. Netsene provides a structurally robust, bulk platform that unifies ultrahigh carrier mobility, topological nodal surfaces, and potential correlation physics, demonstrating the power of theory-guided, machine-learning-accelerated discovery for engineering topological quantum phases.		
	\end{abstract}
	\maketitle
	\section{Introduction}
	\label{sec:intro}
	
	The experimental realization of graphene has established 2D Dirac semimetals as a cornerstone of modern condensed matter physics, unveiling massless relativistic quasiparticles and exceptional carrier mobilities~\cite{Novoselov2004, CastroNeto2009}. A central pursuit in topological materials science is to extend such Dirac physics into 3D, where higher-dimensional band degeneracies—such as one-dimensional (1D) nodal lines and 2D nodal surfaces—can emerge~\cite{Burkov2011, Armitage2018}. Among these, nodal surfaces, across which two bands remain degenerate on an entire 2D manifold in momentum space, are particularly intriguing. They can host extended regions of high density of states, potentially fostering enhanced electronic correlations and exotic transport phenomena~\cite{Wu2018a}. However, realizing clean, robust nodal surfaces near the Fermi level in stable materials remains a significant challenge. Their existence often demands specific, non-symmorphic crystal symmetries and a delicate balance of electronic interactions, making both their rational design and experimental realization difficult~\cite{Bzdusek2017}.
	
	\begin{figure*}
		\centering
		\includegraphics[width=0.95\linewidth]{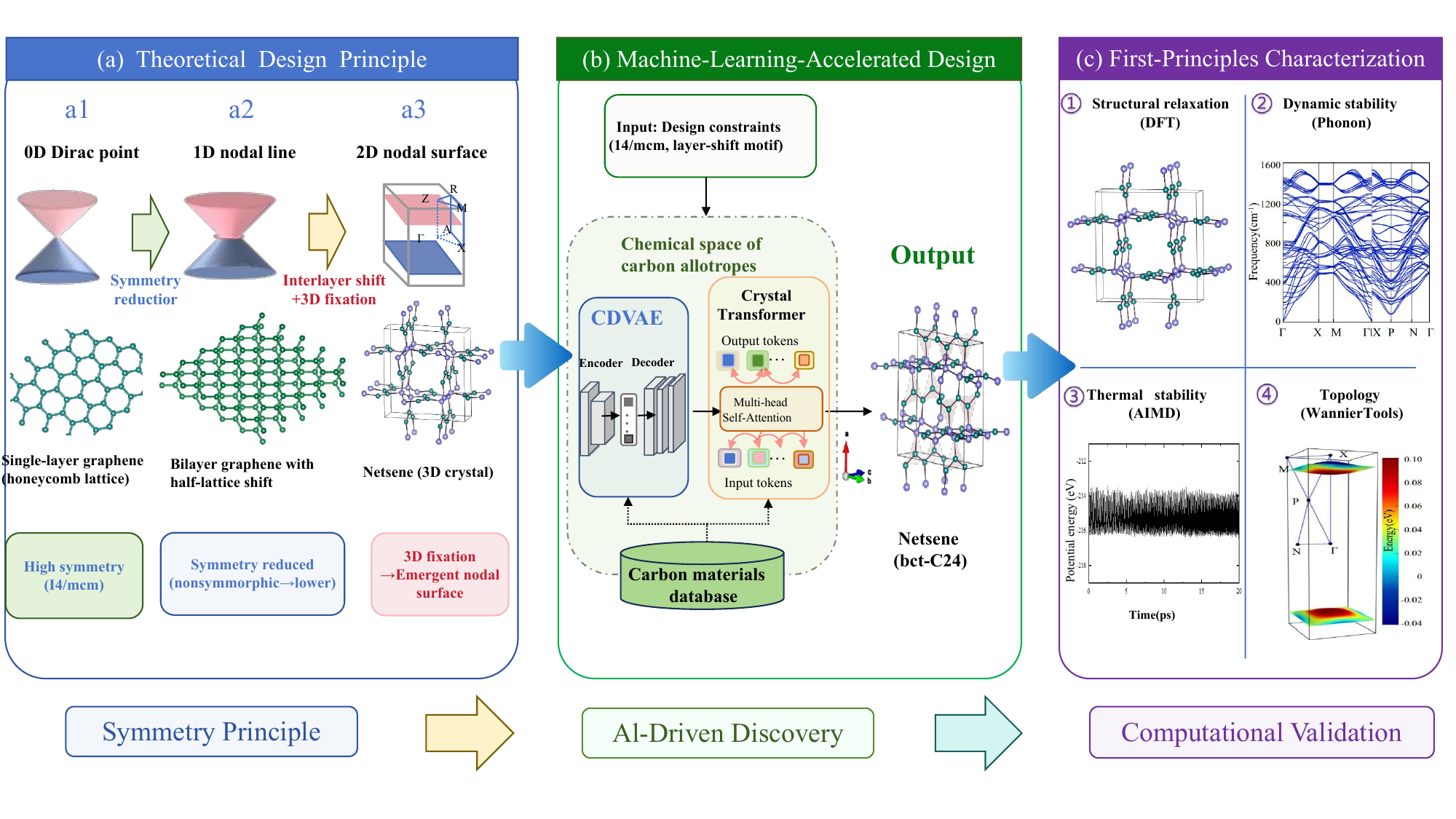}
		\caption{Schematic overview of the integrated research methodology.
			(a) Three-step symmetry-engineering design principle for transforming a 2D Dirac cone into a 3D nodal surface. (b) Workflow of the machine-learning-accelerated inverse design combining a CDVAE and a Crystal Transformer. (c) Key first-principles characterization methods used to validate the predicted structure and its topological properties.}
		\label{fig:method}
	\end{figure*}
	Elemental carbon, with its weak spin-orbit coupling and extraordinary structural diversity ($sp$, $sp^2$, $sp^3$ hybridization), offers an ideal materials platform for engineering topological band structures~\cite{Wang2018, Wang2019}. The weak spin-orbit coupling preserves band degeneracies protected by spacetime inversion symmetry, while carbon's versatile bonding enables the construction of complex 3D networks with tailored symmetries. This raises a fundamental question: \textit{Can graphene's Dirac cones be rationally transformed into symmetry-protected nodal surfaces in a 3D carbon allotrope?}
	
	In this work, we answer this question affirmatively by proposing and validating a general symmetry-engineering principle that systematically elevates the dimensionality of band degeneracies. Our strategy unfolds in three conceptual steps: (i) reducing the in-plane hexagonal symmetry of graphene to break the equivalence of hopping directions, promoting the 0D Dirac point into a 1D nodal line; (ii) introducing a controlled half-lattice-constant registry shift between graphene layers to extend this nodal line; and (iii) permanently ``fixing'' the shifted layers with a covalent $sp^3$ carbon network, which simultaneously stabilizes a body-centered tetragonal structure and activates non-symmorphic symmetries that enforce a 2D nodal surface. Rather than relying on trial-and-error, we translate this design principle into a target for machine-learning-accelerated inverse design. By combining a CDVAE~\cite{Xie2021}  with a Crystal Transformer~\cite{Cao2025}, we efficiently navigate the vast chemical space of carbon allotropes and identify a thermodynamically and dynamically stable phase that perfectly embodies our design.
	
	The discovered material, which we name \textbf{Netsene} (bct-C$_{24}$), crystallizes in the \textit{I4/mcm} space group and exhibits a remarkable set of properties. First-principles calculations reveal that Netsene is not only mechanically robust—with an exceptionally high stiffness along the $c$-axis ($C_{33}=1134.2$ GPa)—but also a unique topological semimetal. Its electronic structure features a complex, \textit{nested nodal-surface system} around the Fermi level, accompanied by Dirac-like linear crossings with Fermi velocities comparable to that of graphene ($\sim 9\times10^5$ m/s). The non-trivial bulk topology manifests in protected drumhead surface states, including a nearly flat band that suggests potential for correlation-driven phenomena. In this regard, Netsene can be viewed as a 3D crystalline analogue of magic-angle TBG, where the momentum-space nodal surface plays a role akin to the Moir\'{e} flat bands, but realized in a structurally robust bulk allotrope. The combination of ultrahigh carrier mobility, topological nodal surfaces, and excellent mechanical stability establishes Netsene as a compelling platform for both fundamental topological physics and potential high-speed electronic applications, and highlights the power of theory-guided, machine-learning-accelerated materials discovery.

\section{Methods}
\label{sec:methods}

\subsection{Theoretical Design Principle}
\label{sec:design}

Our strategy to engineer a 3D nodal-surface semimetal from graphene is a three-step symmetry-engineering procedure that systematically raises the dimensionality of the band degeneracy. We outline each step below with analytical model derivations.

\textit{Step 1: From 0D Dirac point to 1D nodal line via in-plane symmetry reduction.}  
Monolayer graphene has a honeycomb lattice with two atoms per unit cell. In the nearest-neighbor tight-binding approximation, its Hamiltonian in momentum space is
\begin{equation}
	H_0(\mathbf{k}) = 
	\begin{pmatrix}
		0 & t f(\mathbf{k}) \\
		t f^*(\mathbf{k}) & 0
	\end{pmatrix},
	\quad f(\mathbf{k}) = \sum_{j=1}^3 e^{i\mathbf{k}\cdot\mathbf{\delta}_j},
	\label{eq:graphene_TB}
\end{equation}
where $t$ is the hopping integral and $\mathbf{\delta}_j$ are the three nearest-neighbor vectors. Expanding around the K point $\mathbf{K} = (4\pi/3a,0)$ yields the massless Dirac Hamiltonian
\begin{equation}
	H_{\text{Dirac}} = v_F (k_x \sigma_x + k_y \sigma_y),
	\label{eq:dirac_2d}
\end{equation}
with Fermi velocity $v_F = \sqrt{3}ta/2$. This is a zero-dimensional point degeneracy.  
If we preserve the honeycomb connectivity but lower the in-plane symmetry from hexagonal to orthorhombic (space group $p2/m$), the equivalence between the three hopping directions is lifted. The effective Hamiltonian near the former K point becomes
\begin{equation}
	H_{\text{ortho}}(\mathbf{q}) = v_x q_x \sigma_x + v_y q_y \sigma_y,
	\label{eq:ortho_Dirac}
\end{equation}
where $v_x \neq v_y$ in general. In an isolated 2D system this remains a point node, but when interlayer coupling is introduced in the next step, the reduced symmetry can promote it into a line degeneracy.

\textit{Step 2: From 1D nodal line to 2D nodal surface via interlayer registry shift.}  
Consider two such graphene layers stacked with a relative shift $\mathbf{d} = (a/2, 0)$ along the armchair direction. The interlayer coupling is modeled by a hopping matrix between the layers. In the basis of the two layers (each with A/B sublattices), the total bilayer Hamiltonian is
\begin{equation}
	H_{\text{bilayer}}(\mathbf{k}) = 
	\begin{pmatrix}
		H_0(\mathbf{k}) & T(\mathbf{k}) \\
		T^\dagger(\mathbf{k}) & H_0(\mathbf{k})
	\end{pmatrix},
	\label{eq:bilayer}
\end{equation}
where $T(\mathbf{k})$ is the interlayer hopping matrix. For the armchair shift $(a/2,0)$, the diagonal (sublattice-preserving) interlayer hopping becomes momentum-dependent and can vanish along certain lines in the Brillouin zone, generating a protected nodal line. The low-energy effective model around such a nodal line is
\begin{equation}
	H_{\text{line}}(\mathbf{q}) = v_\perp (q_1 \sigma_x + q_2 \sigma_y) + w(q_1,q_2)\sigma_0,
	\label{eq:nodalline_model}
\end{equation}
where $q_1,q_2$ are momenta perpendicular to the line, and $w$ tilts the cone. The line itself is protected by a $\mathbb{Z}_2$ invariant (Berry phase $\pi$) when time-reversal and sublattice symmetries are present.

\textit{Step 3: Permanent 3D fixation and symmetry-protected nodal surface.}  
To extend the nodal line into a 2D nodal surface, we permanently connect the two shifted layers with an additional carbon network that forms a square lattice. This covalently ``fixes'' the shifted geometry into a bulk structure. The resulting crystal is body-centered tetragonal (bct) with space group \textit{I4/mcm} (No.~140). This space group possesses non-symmorphic symmetry operations, notably a screw axis $\{C_{2z} | (\frac{1}{2},\frac{1}{2},0)\}$ and a glide mirror $\{M_y | (\frac{1}{2},\frac{1}{2},0)\}$, which are essential for stabilizing the nodal surface.  
In the 3D periodic structure, the interlayer coupling becomes $k_z$-dependent. The effective Hamiltonian near the former nodal line can be written as
\begin{equation}
	H_{\text{3D}}(\mathbf{k}) = 
	\begin{pmatrix}
		\varepsilon_0(\mathbf{k}) & \Delta(\mathbf{k}) \\
		\Delta^*(\mathbf{k}) & \varepsilon_0(\mathbf{k})
	\end{pmatrix},
	\label{eq:effective_3D}
\end{equation}
where $\Delta(\mathbf{k})$ is the effective hybridization between the two shifted graphene subunits. Owing to the bct symmetry constraints, $\Delta(\mathbf{k})$ is forced to vanish on a 2D manifold in the Brillouin zone,
\begin{equation}
	\Delta(\mathbf{k}) = 0 \quad \Rightarrow \quad \text{nodal surface } \mathcal{S}.
	\label{eq:nodal_surface_condition}
\end{equation}
This surface is not accidental but symmetry-enforced, and the degeneracy is topologically protected by a $\mathbb{Z}_2$ invariant derived from the Berry phase of loops encircling the surface~\cite{Wu2018a}. The idea of chemically fixing shifted graphene layers into a stable 3D network is reminiscent of hydrogenated graphene (graphane), where $sp^2$ to $sp^3$ rehybridization leads to new electronic phases~\cite{Sofo2007}. Recent experimental progress in fabricating graphene-on-diamond heterostructures~\cite{Yuan2024} further supports the synthetic plausibility of such a phase. Thus, the search for a 3D nodal-surface semimetal reduces to identifying a stable carbon allotrope that realizes the symmetry conditions of the \textit{I4/mcm} bct structure.

\begin{figure}[tbp]
	\centering
	\includegraphics[width=0.95\linewidth]{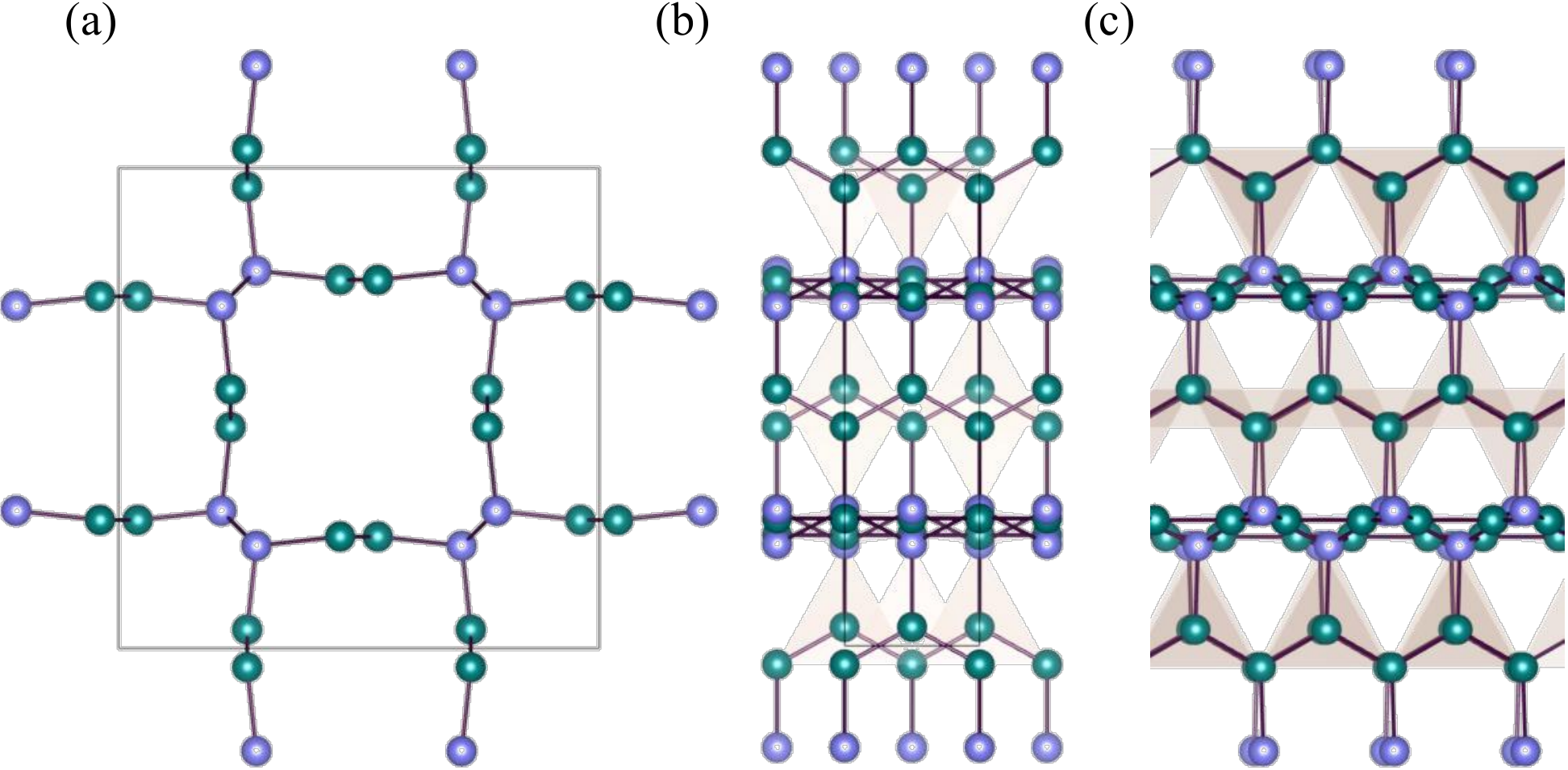}
	\caption{(a) The conventional unit cell of \textbf{Netsene} viewed along the $c$-axis, showing the layered arrangement. C1 ($sp^2$-like) and C2 ($sp^3$-like) atoms are shown in blue and orange, respectively. (b) Side view along the $a$-axis, highlighting the stacking sequence of four distinct layers. (c) Projection along the $a$-axis, clearly showing the armchair-direction shift between adjacent layers. The (110) plane forms the distorted honeycomb lattice.}
	\label{fig:structure}
\end{figure}
\subsection{Machine-Learning-Accelerated Inverse Design}
\label{sec:ml}

The design principle derived above provides a clear target for materials discovery. To efficiently explore the vast chemical space of 3D carbon allotropes, we developed a deep generative framework for crystal structure prediction. The model integrates a CDVAE with a Crystal Transformer architecture~\cite{Cao2025}. The CDVAE generates candidate structures by learning the underlying distribution of stable carbon materials, while the Transformer-based property predictor guides the search toward candidates that satisfy the required bct symmetry and structural motifs (layered, half-shifted graphene subunits). Our training dataset was assembled from the Materials Project~\cite{Jain2013}, the Samara Carbon Allotrope Database~\cite{Hoffmann2016}, and a comprehensive set of theoretically proposed 3D carbon networks from recent literature. The generative process was constrained to explore structures compatible with the \textit{I4/mcm} space group and the layer-shift design, rapidly leading to a novel, low-energy candidate: \textbf{Netsene} (bct-C$_{24}$).

\subsection{First-Principles Calculations}
\label{sec:df}

All first-principles calculations in this work were performed using the Quantum ESPRESSO(QE) suite~\cite{Giannozzi2017}. The electronic exchange-correlation interactions were treated within the generalized gradient approximation (GGA) as parameterized by the Perdew-Burke-Ernzerhof (PBE) functional~\cite{Perdew1996}. Optimized norm-conserving Vanderbilt pseudopotentials were employed with a kinetic energy cutoff of 80 Ry for the wavefunctions and 600 Ry for the charge density, ensuring excellent convergence for the mixed $sp^2$--$sp^3$ carbon networks. For the pristine bulk unit cell of Netsene, a $\Gamma$-centered Monkhorst-Pack \textbf{6~$\times$~6~$\times$~20} k-point mesh was used for Brillouin zone sampling. Structural relaxations were carried out by fully optimizing both atomic coordinates and lattice vectors until the total energy change between successive steps was below \textbf{10$^{-8}$~Ry}, and forces on each atom were less than $10^{-6}$ Ry/Bohr.

Phonon dispersions were calculated using density functional perturbation theory (DFPT) as implemented in QE to assess dynamic stability. \textit{Ab initio} molecular dynamics (AIMD) simulations were performed at 500~K for 5~ps with a time step of 1~fs to verify thermal stability. The topological surface states and Fermi arcs were calculated using the surface Green's function method as implemented in the WannierTools package~\cite{Wu2018}. Maximally localized Wannier functions (MLWFs) for the carbon $p_z$ orbitals were constructed using the Wannier90 code~\cite{Mostofi2008}, yielding a faithful tight-binding Hamiltonian for the surface state calculations.
	
	\section{Results and Discussion}
	\label{sec:results}

	\subsection{Crystal Structure and Stability of Netsene}
	\label{sec:structure}
	
	Guided by the design principle and identified through our ML-accelerated search, the predicted carbon allotrope, Netsene, crystallizes in a body-centered tetragonal (bct) lattice with space group \textit{I4/mcm} (No.~140). Its optimized equilibrium lattice parameters are $a = b = 8.6913$\,\AA\ and $c = 2.4674$\,\AA. The conventional unit cell contains 24 carbon atoms occupying two inequivalent Wyckoff positions: C1 at $4c$ $(0.03926, 0.73720, 0.5)$ and C2 at $16l$ $(0.71365, 0.78635, 0.5)$. The crystal structure is illustrated in Fig.~\ref{fig:structure}.
	
	As shown in Figs.~\ref{fig:structure}(a) and \ref{fig:structure}(b), the structure faithfully realizes our design: it consists of graphene-like layers stacked along the $c$-axis with a specific half-lattice-constant shift along the armchair direction, clearly visualized in the $a$-axis projection of Fig.~\ref{fig:structure}(c). This stacking pattern generates a 3D covalent network in which C1 atoms primarily exhibit $sp^2$-like hybridization, forming distorted hexagonal rings within the layers, while C2 atoms adopt an $sp^3$-like hybridization, serving as interlayer connectors. This mixed hybridization is key to the structure's 3D rigidity while preserving the $\pi$-electron network essential for the topological electronic structure.
	
	The distorted honeycomb lattice in the (110) plane features three distinct bond lengths: C1--C1~=~1.410\,\AA, C1--C2~=~1.525\,\AA, and C2--C2~=~1.523\,\AA. Notably, the C1--C1 bond is slightly shorter than the C--C bond in graphene (1.42\,\AA), indicating in-plane bond strengthening, while the interlayer bonds (C1--C2, C2--C2) are considerably shorter than the $sp^3$ bond in diamond (1.55\,\AA), reflecting a highly cohesive 3D network.
	
\begin{figure}[t]
		\centering
		\includegraphics[width=0.95\linewidth]{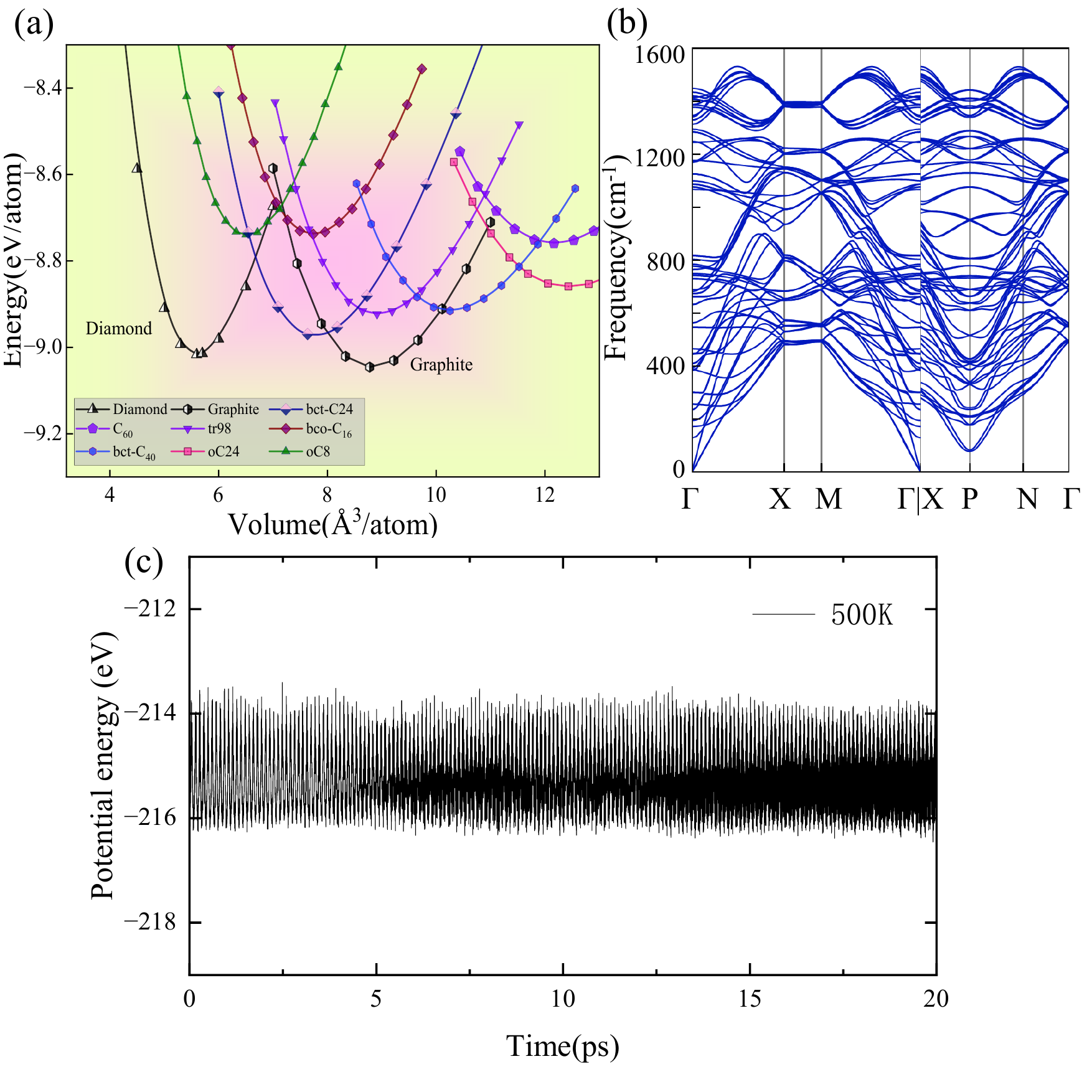} 
		\caption{(a) The energy–volume curve of \textbf{Netsene}, compared with diamond, graphite, and the previously reported oC8, bco-C16, oC24, bct-C40, C60, and tr98 structures. (b) Phonon band structures of \textbf{Netsene} under zero pressure. The highest phonon frequency is approximately 1532 cm$^{-1}$, which is close to the highest phonon frequency of graphite (1610 cm$^{-1}$)\cite{Maultzsch2004}. (c) Potential energy fluctuations during AIMD simulations at 500 K, confirming thermal stability of \textbf{Netsene}.}
		\label{fig:stability}
	\end{figure}

	We rigorously assessed the stability of Netsene through multiple complementary approaches. The calculated energy-volume curve per atom is shown in Fig.~\ref{fig:stability}(a), alongside those of diamond, graphite, and several previously reported carbon allotropes. Netsene is a metastable phase: it lies higher in energy than diamond and graphite, but is notably more stable than other theoretically predicted phases such as tr98, bct-C40, oC24, C60, bco-C16, and oC8~\cite{Li2015,Wang2016,Li2018,Wang2018,Dav1991,Wang2019}. This energetic ranking places Netsene among the most favorable low-energy metastable carbon allotropes predicted to date.
	
	Phonon dispersion calculations, performed across the entire Brillouin zone using DFPT, are presented in Fig.~\ref{fig:stability}(b). The absence of any imaginary frequency modes confirms the dynamic stability of Netsene, indicating that the structure resides at a genuine local minimum on the potential energy surface. The highest phonon frequency reaches approximately 1532~cm$^{-1}$, which is close to that of graphite (1610~cm$^{-1}$)~\cite{Maultzsch2004} and indicative of strong covalent bonding. Furthermore, \textit{ab initio} molecular dynamics (AIMD) simulations at 500~K [Fig.~\ref{fig:stability}(c)] demonstrate excellent thermal stability: the structure remains intact with negligible fluctuations in total potential energy and no bond-breaking events throughout the 5~ps simulation, confirming that Netsene can maintain its structural integrity well above room temperature.

	\subsection{Mechanical Properties}
	\label{sec:mechanics}
	
	The mechanical robustness of Netsene was quantified by calculating its full elastic constant tensor $C_{ij}$. For a tetragonal crystal (point group $4/mmm$), there are six independent elastic constants: $C_{11}$, $C_{12}$, $C_{13}$, $C_{33}$, $C_{44}$, and $C_{66}$. Our calculated values are listed in Table~\ref{tab:elastic}, alongside those of diamond and graphite for comparison. All $C_{ij}$ values satisfy the generalized Born-Huang stability criteria for a tetragonal system:
	\begin{equation}
		C_{11} > |C_{12}|,\; C_{44} > 0,\; C_{66} > 0,\; 2C_{13}^2 < C_{33}(C_{11}+C_{12}),
		\label{eq:born}
	\end{equation}
	further corroborating its mechanical stability.
	
	\begin{table}[htbp]
		\centering
		\caption{Calculated elastic constants $C_{ij}$ (in GPa), bulk modulus $B$, shear modulus $G$, Young's modulus $E$ (in GPa), and Poisson's ratio $\nu$ for Netsene, compared with diamond and graphite. All values are from PBE-DFT calculations for consistency.}
		\label{tab:elastic}
		\begin{tabular}{lccccccccc}
			\toprule
			\textbf{Material} & $C_{11}$ & $C_{33}$ & $C_{44}$ & $C_{66}$ & $C_{12}$ & $C_{13}$ & $B$ & $G$ & $E$ \\
			\midrule
			Netsene & 551.1 & 1134.2 & 278.7 & 70.1 & 98.2 & 165.4 & 290 & 243 & 570 \\
			Diamond~\cite{McSkimin1972} & 1079 & 1079 & 578 & 578 & 124 & 124 & 451 & 535 & 1143 \\
			Graphite~\cite{Bosak2007} & 1109 & 38.7 & 4.95 & 4.5 & 139 & 0 & 280 & 4.2 & 12 \\
			\bottomrule
		\end{tabular}
	\end{table}
	
	The derived bulk modulus $B$ of Netsene is 290~GPa, comparable to that of graphite (280~GPa) but lower than diamond's 451~GPa, indicating substantial incompressibility for a low-density carbon framework. The shear modulus $G$ (243~GPa) and Young's modulus $E$ (570~GPa) are remarkably high, surpassing those of most known 3D carbon allotropes and many advanced ceramics. This places Netsene in the category of ultra-strong, lightweight materials.
	
	Most strikingly, the value of $C_{33} = 1134.2$~GPa is exceptionally high, exceeding even the in-plane elastic constant of graphene ($C_{11} \sim 1100$~GPa). This reveals that Netsene possesses outstanding stiffness specifically along the crystallographic $c$-axis—the stacking direction. This unique anisotropy arises directly from the structural motif: the $sp^3$-bonded C2 connectors form nearly linear, highly rigid chains parallel to the $c$-axis, effectively bridging the $sp^2$ honeycomb layers. Netsene thus combines the in-plane strength characteristic of graphene with an unprecedented out-of-plane rigidity, a combination that is highly desirable for structural applications and that directly reflects the success of our ``fixing'' design strategy.
	
	\subsection{Electronic Structure and the Nested Nodal-Surface System}
	\label{sec:electronic}
	
\begin{figure}[t]
		\centering
		\includegraphics[width=\linewidth]{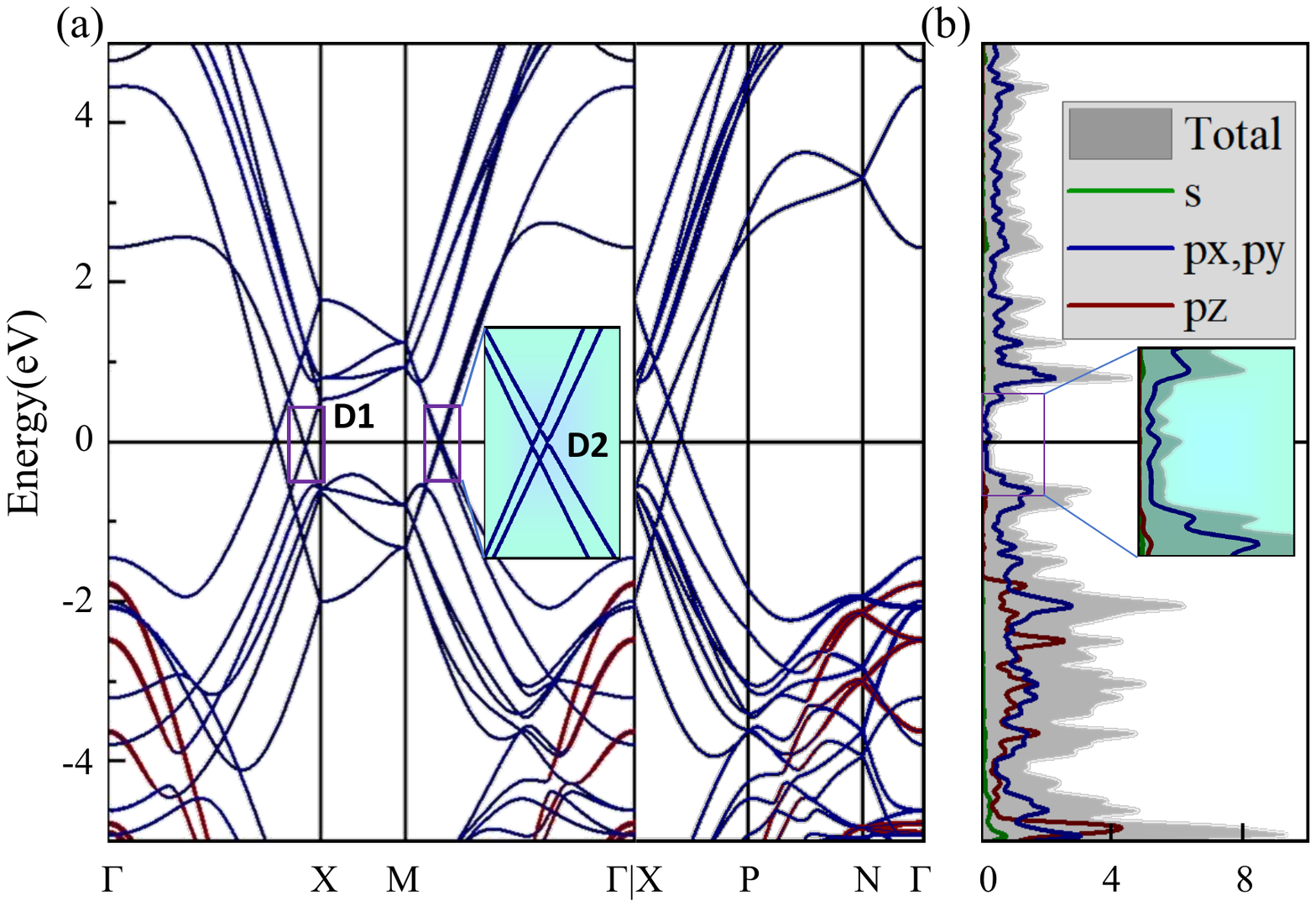}
		\caption{(a) Electronic band structure of Netsene along high-symmetry lines. The inset zooms in on the linear crossings near D$_2$. (b) Total and atom-projected density of states (DOS and PDOS). The Fermi level is set to zero.}
		\label{fig:band}
	\end{figure}

	We now turn to the defining electronic feature of Netsene: its topological nodal-surface semimetal character. The electronic band structure along high-symmetry paths, calculated without spin-orbit coupling (SOC), is presented in Fig.~\ref{fig:band}(a). Netsene exhibits a clear semimetallic character, with multiple bands crossing the Fermi level ($E_F$). Two prominent linear band crossings are observed near $E_F$: one along the $\Gamma$--X path (denoted D$_1$), located at $k \approx 1.7953$\,\AA$^{-1}$ and approximately 80~meV below $E_F$, and another near the M--$\Gamma$ region (denoted D$_2$), located at $k \approx 4.2928$\,\AA$^{-1}$. 
	
	A detailed linear fit to these crossings confirms their massless Dirac-fermion nature. The Fermi velocities extracted for D$_1$ are $v_F^{\text{val}} \approx 9.03 \times 10^5$~m/s and $v_F^{\text{con}} \approx 8.37 \times 10^5$~m/s, while those for D$_2$ are $v_F^{\text{val}} \approx 9.05 \times 10^5$~m/s and $v_F^{\text{con}} \approx 9.99 \times 10^5$~m/s. These values are on the same order of magnitude as the Fermi velocity of pristine graphene ($\sim 8.5 \times 10^5$~m/s)~\cite{Deacon2007}, suggesting that Netsene inherits graphene's ultrahigh carrier mobility and extends it into a 3D bulk crystal.
	
	The overall density of states (DOS) at $E_F$ is finite but relatively low, characteristic of a semimetal [Fig.~\ref{fig:band}(b)]. The projected DOS (PDOS) reveals that the states near $E_F$ are predominantly contributed by the $p$ orbitals of both C1 and C2 atoms, confirming that the low-energy physics is governed by the $\pi$-bonding network inherited from the graphene-like layers.
	
\begin{figure}[t]
		\centering
		\includegraphics[width=0.95\linewidth]{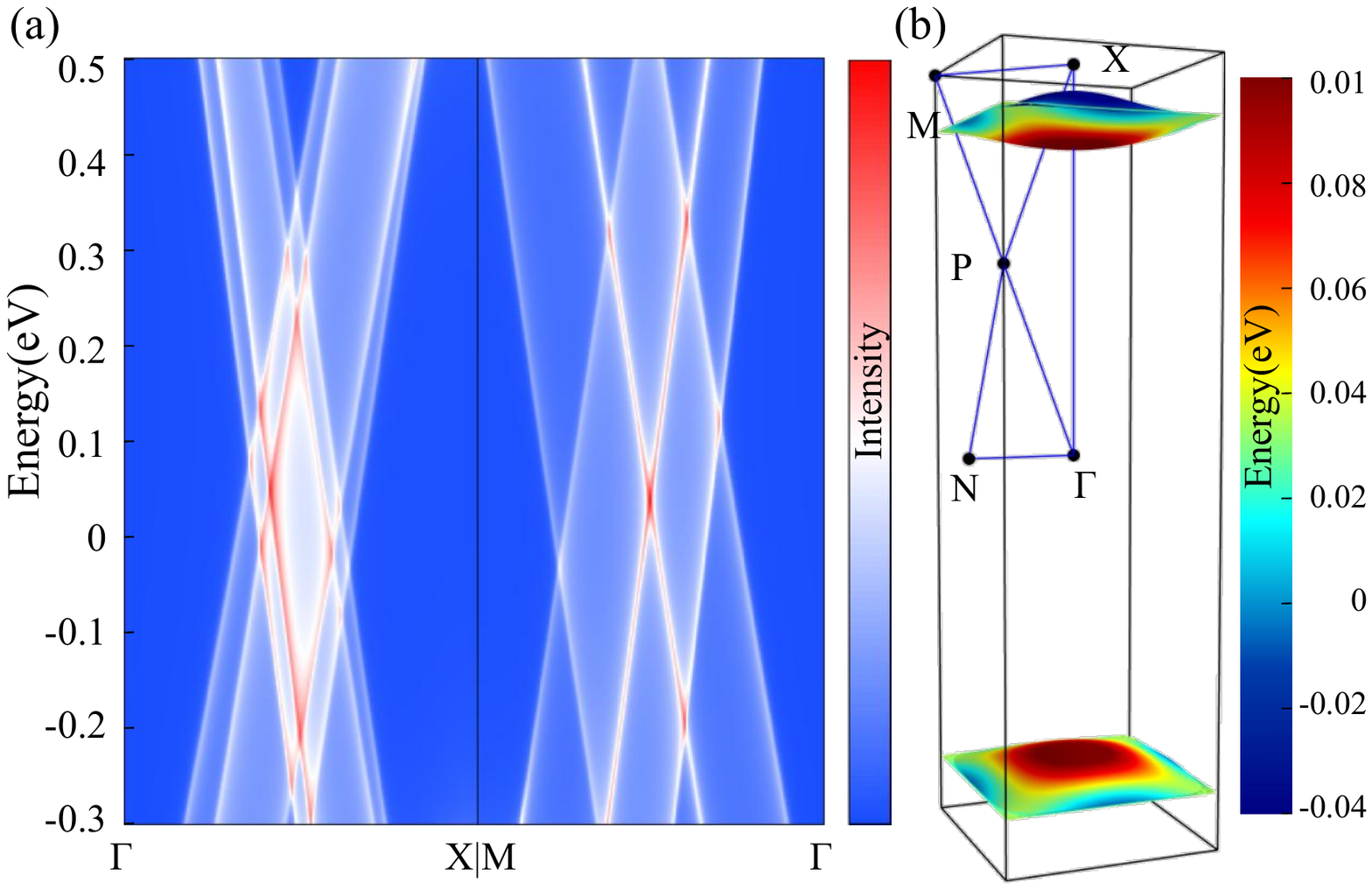}
		\caption{(a) 3D visualization of the nested nodal-surface system around the Fermi level, showing the ``double-bowl'' shaped degeneracy manifolds. (b) Schematic of the constant-$k_z$ evolution of the band crossings, illustrating how the nodal surfaces form and annihilate as a function of $k_z$.}
		\label{fig:nodal}
	\end{figure}

	Beyond these isolated Dirac-like points, the most striking feature of Netsene's electronic structure is the presence of a complex, extended degeneracy forming a \textbf{nested nodal-surface system}. This is visualized in the 3D band structure plot of Fig.~\ref{fig:nodal}(a), where a series of nearly flat bands intersect and overlap near $E_F$, creating 2D manifolds of band degeneracy in specific regions of the Brillouin zone.
	
	Specifically, we identify four distinct nodal surfaces centered around $k_z \approx \pm 0.42\,(2\pi/c)$. The genesis and evolution of these degeneracies are elucidated by examining constant-$k_z$ slices of the band structure. At $k_z = 0.39\,(2\pi/c)$, near the $\Gamma$ point ($k_x=k_y=0$), two pairs of bands approach each other, forming two separate linear crossings. As $k_z$ increases to $0.42\,(2\pi/c)$, these four bands converge to a four-fold degenerate point precisely at $\Gamma$. Simultaneously, a separate four-fold degeneracy emerges at the zone boundary ($k_x=k_y=0.5$). Upon further increasing $k_z$ to $0.46\,(2\pi/c)$, the degeneracies split again into two Dirac-like points. This evolution, illustrated schematically in Fig.~\ref{fig:nodal}(b), creates a distinctive nested, ``double-bowl'' shaped nodal surface in the 3D Brillouin zone.
	
	The origin of this robust nodal system is directly traceable to the non-symmorphic space group symmetry of Netsene (\textit{I4/mcm}) and the specific half-lattice-constant shift between layers dictated by our design principle. The screw axis $\{C_{2z} | (\frac{1}{2},\frac{1}{2},0)\}$ and glide mirror $\{M_y | (\frac{1}{2},\frac{1}{2},0)\}$ protect the band crossings along high-symmetry lines and enforce the formation of extended nodal surfaces rather than isolated lines or points. The $\pi$-orbital hybridization pattern, modulated by the shifted stacking and the $sp^3$ connectors, generates precisely the number of closely spaced bands necessary for this complex degeneracy. A non-trivial $\mathbb{Z}_2$ topological invariant, evaluated from the Berry phase on a loop encircling any of these nodal surfaces, guarantees their topological protection~\cite{Wu2018a}.
	
	A direct spectroscopic signature of this nested nodal-surface system appears in the electronic DOS. As seen in Fig.~\ref{fig:band}(b), a pronounced peak emerges just above $E_F$. This peak corresponds to a van Hove singularity arising from the extended regions of nearly flat dispersion associated with the nodal surfaces. This feature is highly reminiscent of the physics of TBG, where flat bands and van Hove singularities arise from Moir\'{e} superlattice potentials~\cite{Cao2018, Bistritzer2011}. In a profound sense, Netsene can be viewed as a \textbf{3D crystalline analogue of magic-angle TBG}: whereas TBG uses a rotational misorientation to generate real-space Moir\'{e} flat bands, Netsene employs a translational layer shift permanently fixed by covalent bonds to generate momentum-space nodal surfaces with an attendant high density of states. This analogy positions Netsene as a structurally robust, bulk platform for exploring correlation-driven phenomena that are actively pursued in 2D Moir\'{e} systems, but with the advantages of three-dimensionality and intrinsic stability.
	
	\subsection{Topological Surface States}
	\label{sec:surface}
	
\begin{figure}[htbp]
		\centering
		\includegraphics[width=0.95\linewidth]{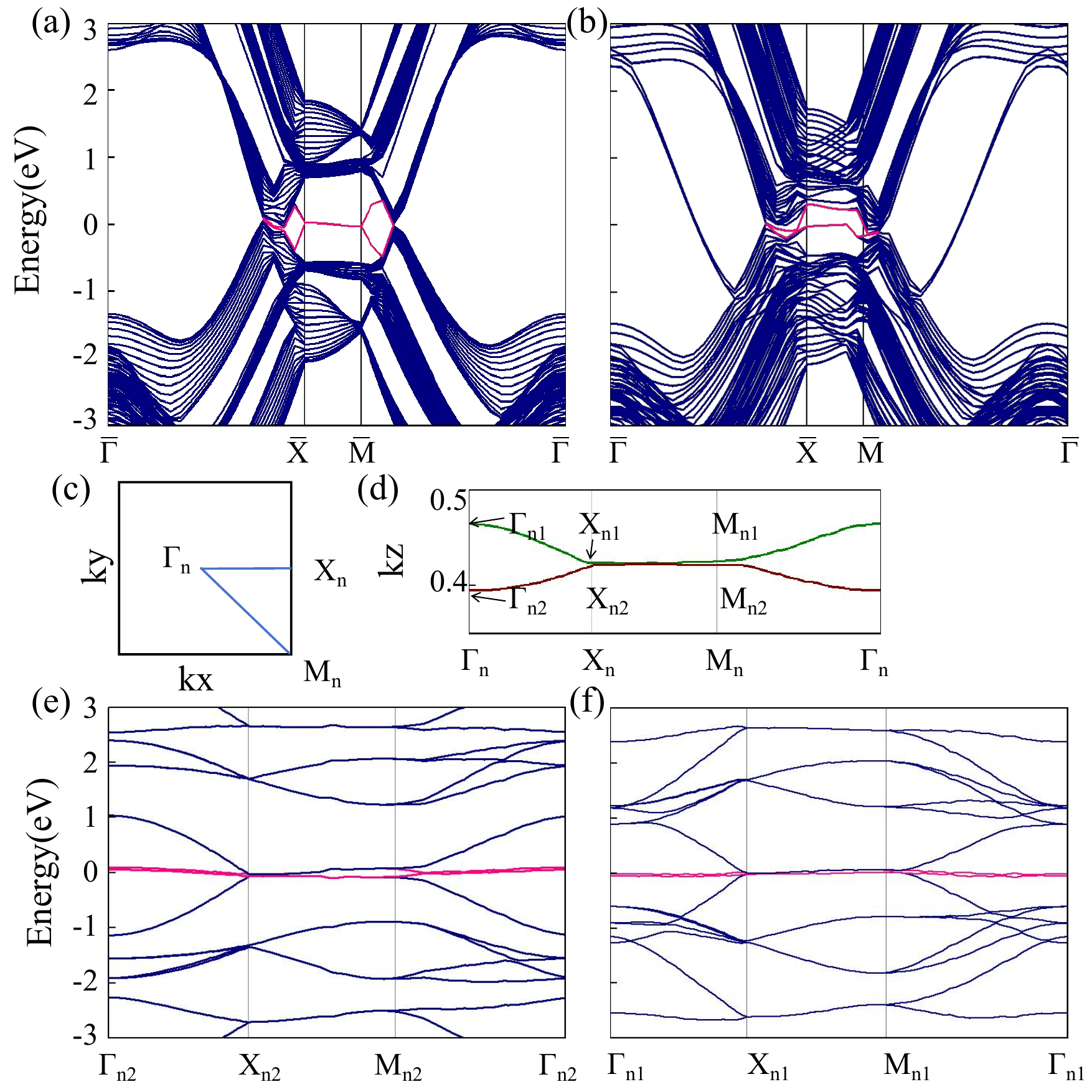}
		\caption{Topological surface states of Netsene. (a) (110) surface spectral function and (b) (100) surface spectral function from a ten-layer slab calculation. Pink dotted lines indicate the top and bottom surface bands, which are degenerate due to inversion symmetry. (c) High-symmetry path in the projected 2D Brillouin zone. (d) Band dispersion showing doubly degenerate bands crossing at $X_n$ and $M_n$. (e),(f) Projected nodal-line band structures confirming continuous linear crossings and drumhead surface states.}
		\label{fig:surface}
	\end{figure}

	The non-trivial bulk topology of Netsene dictates the existence of topological surface states. Figures~\ref{fig:surface}(a) and \ref{fig:surface}(b) show the calculated surface spectral functions for the (110) and (100) surfaces, respectively, obtained from a ten-layer-thick slab geometry with surface dangling bonds saturated by hydrogen atoms. Due to the inversion-symmetric slab model employed, the top and bottom surface bands are degenerate and are indicated by pink dotted lines in both panels.
	
	The surface bands exhibit nearly dispersionless flat features and are concentrated in the interior region of the $\bar{X}$–$\bar{M}$ path, i.e., on the side of the projected nodal surface containing the $\Gamma$ point.  Such a distribution pattern is distinctly different from that of conventional nodal‑line semimetals, in which drumhead surface states typically emerge only at Brillouin‑zone boundaries (i.e., the $\bar{X}$ or $\bar{M}$ points).This flat dispersion is a typical signature of drumhead surface states in nodal‑surface semimetals.
	
	To further confirm the topological origin of these surface features, we analyze the projected nodal-line band structure. Figure~\ref{fig:surface}(c) defines the high-symmetry path in the projected 2D Brillouin zone, along which the band dispersion is calculated. The results in Fig.~\ref{fig:surface}(d) reveal a pair of doubly degenerate bands spanning energies from approximately 0.4 to 0.5~eV. These bands exhibit pronounced degeneracies at the high-symmetry points $X_n$ and $M_n$ and undergo direct band crossings at $X_n$—a hallmark signature of topological nodal lines projected onto the surface. The continuity and topological stability of these nodal lines are further verified in Figs.~\ref{fig:surface}(e) and \ref{fig:surface}(f), which display the projected nodal-line band structures. Continuous linear crossings are observed along the paths $\Gamma_{n2}\to X_{n2}\to M_{n2}\to \Gamma_{n2}$ and $\Gamma_{n1}\to X_{n1}\to M_{n1}\to \Gamma_{n1}$, with the crossing points fully consistent with the degenerate positions in Fig.~\ref{fig:surface}(d). These drumhead surface states, and especially the nearly flat surface band, are a direct consequence of the bulk nested nodal-surface topology and constitute a readily testable experimental signature for ARPES measurements.
	
	\section{Summary}
	\label{sec:summary}
	
	In summary, we have proposed and validated a general symmetry-engineering principle that systematically transforms the Dirac cone of 2D graphene into a 3D nested nodal-surface semimetal. Guided by this principle and accelerated by a deep generative model, we discovered Netsene (bct-C$_{24}$), a novel, dynamically and mechanically stable carbon allotrope. Netsene is a unique topological semimetal hosting a complex nested nodal-surface system around the Fermi level, accompanied by Dirac-like linear crossings with Fermi velocities comparable to graphene. Its non-trivial bulk topology manifests in drumhead surface states, including a nearly flat surface band. The combination of ultrahigh carrier mobility, topological nodal surfaces, high mechanical stiffness ($C_{33}=1134.2$~GPa), and robust stability positions Netsene as a compelling 3D carbon platform. As a crystalline analogue of TBG, Netsene offers a structurally robust arena for exploring correlation physics and high-speed electronic applications in a bulk material. This work demonstrates the power of theory-guided, machine-learning-accelerated materials discovery for engineering exotic topological quantum phases.
	 
	$Acknowledgments.$ We wish to thank K. Bu and D.-X. Yao for helpful discussions. This work is supported by Beijing National Laboratory for Condensed Matter Physics (Grant No. 2024BNLCMPKF020), Innovation Capability Improvement Project of Hebei province (Grant No. 22567605H), the National Natural Science Foundation of China (Grants No. 12204400).
	
	\bibliographystyle{apsrev4-2}
	\nocite{*}
	\bibliography{c24}
\end{document}